\documentclass[amsmath,amssymb, aps,prapplied, twocolumn,superscriptaddress]{revtex4-2}
\usepackage[utf8]{inputenc}
\usepackage{hyperref}
\hypersetup{
	colorlinks=true,
	linkcolor=blue,
	filecolor=blue,
	citecolor = blue,      
	urlcolor=cyan,
}

\usepackage{amsmath}
\usepackage{amsfonts}
\usepackage{MnSymbol}
\usepackage{xcolor}
\usepackage{graphicx}
\usepackage{siunitx}

\newcommand*{\bra}[1]{\ensuremath{\langle #1 \vert}}
\newcommand*{\ket}[1]{\ensuremath{\vert #1 \rangle}}

\newcommand{\lr}[1]{\left( #1 \right)}
\renewcommand{\vec}[1]{{\boldsymbol{#1}}}

\newcommand{\figref}[1]{Fig.~\ref{#1}}

\newcommand{\secref}[1]{Sec.~\ref{#1}}
\newcommand{\eqeqref}[1]{Eq.~(\ref{#1})}

\newcommand{\mr}[1]{{\mathrm{#1}}}

\usepackage{soul}

\usepackage{orcidlink}

\begin{document}
\title{Sparse Graph Optimization Using Weighted Quantum Wires in Rydberg Atom Arrays}

\author{A. G. de Oliveira\,\orcidlink{0000-0002-0220-8783}}
\email{andre.oliveira@strath.ac.uk}
\affiliation{Department of Physics and SUPA, University of Strathclyde, Glasgow G4 0NG, UK}

\author{J. Kombe\,\orcidlink{0000-0002-9982-2068}}
\affiliation{Department of Physics and SUPA, University of Strathclyde, Glasgow G4 0NG, UK}

\author{G. Pelegr\'i\,\orcidlink{0000-0001-7107-4398}}
\affiliation{Department of Physics and SUPA, University of Strathclyde, Glasgow G4 0NG, UK}

\author{P. Schroff\,\,\orcidlink{0009-0006-0943-8318}}
\affiliation{Department of Physics and SUPA, University of Strathclyde, Glasgow G4 0NG, UK}

\author{M. T. Wells-Pestell\,\orcidlink{0009-0009-3018-414X}}
\affiliation{Department of Physics and SUPA, University of Strathclyde, Glasgow G4 0NG, UK}

\author{D. M. Walker\,\orcidlink{0000-0002-4206-2942}}
\affiliation{Department of Physics and SUPA, University of Strathclyde, Glasgow G4 0NG, UK}

\author{A. J. Daley\,\orcidlink{0000-0001-9005-7761}}
\affiliation{Department of Physics and SUPA, University of Strathclyde, Glasgow G4 0NG, UK}
\affiliation{Department of Physics, University of Oxford, Clarendon Laboratory, OX1 3PU Oxford, UK}

\author{J. D. Pritchard\,\orcidlink{0000-0003-2172-7340}}
\email{jonathan.pritchard@strath.ac.uk}
\affiliation{Department of Physics and SUPA, University of Strathclyde, Glasgow G4 0NG, UK}

\begin{abstract}
Neutral atom arrays provide a versatile platform to implement coherent quantum annealing as an approach to solving hard combinatorial optimization problems. Here we present and experimentally demonstrate an efficient encoding scheme based on chains of Rydberg-blockaded atoms, which we call quantum wires, to natively embed maximum weighted independent set (MWIS) and quadratic unconstrained binary optimization (QUBO) problems on a neutral atom architecture. For graphs with quasi-unit-disk connectivity, in which only a few long-range edges are required, our approach requires a significantly lower overhead in the number of ancilla qubits than previous proposals, facilitating the implementation on currently available hardware. To demonstrate the approach, we perform  annealing of weighted graphs on a programmable atom array using local light-shifts to encode problem-specific weights across graphs of varying sizes. This approach successfully identifies the solutions to the original MWIS and QUBO graph instances. Our work expands the operational toolkit of near-term neutral atom arrays, enhancing their potential for scalable quantum optimization.
\end{abstract}
\maketitle

\section{Introduction}\label{sec:intro}
    Combinatorial optimization problems are of great relevance in a wide range of scientific fields and industry sectors \cite{Paschos2014}. In many cases, these problems have been shown to be NP-hard, and finding new algorithms to tackle them more efficiently remains an outstanding challenge. Over the last two decades, quantum optimization has attracted attention as a potential route towards achieving computational speedups for certain problems \cite{abbas2024challenges}. Quantum optimization approaches typically rely on preparing a many-body quantum system in a ground state that encodes the solution of the combinatorial optimization problem at hand. Typically, this state preparation is achieved through quantum annealing algorithms (QAA) \cite{HaukeOliver2020,RajakChakrabarti2023} that rely on the quantum adiabatic principle, or using hybrid quantum-classical approaches \cite{CerezoColes2021,AstrakhantsevCarleo2023} such as the Variational Quantum Eigensolver (VQE) \cite{PeruzzoOBrien2014,KandalaGambetta2017,HempelRoos2018,KokailZoller2019}. These approaches are particularly well suited for combinatorial optimization problems that admit formulations as spin models, with examples including maximum (weighted) independent set (MIS/MWIS) and quadratic unconstrained binary optimization (QUBO) problems \cite{lucas2014ising}.
 
\begin{figure}[t!]
  \includegraphics[width=\linewidth]{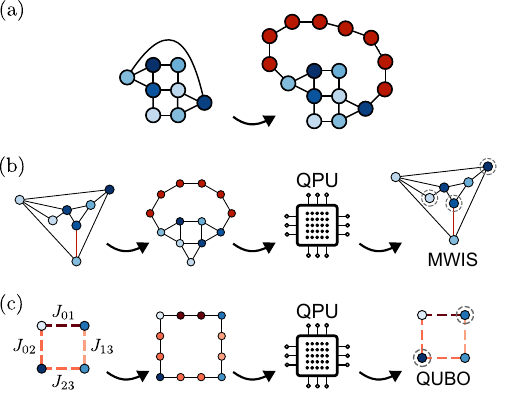} 
    \vspace{-1cm}
    \caption{{\bf Quantum Wire Approach} (a) An example MWIS problem defined on a graph with non-unit-disk connectivity imposed by a sparse number of long-range edges (left) can be efficiently embedded on neutral atom hardware by introducing weighted quantum wires (right) created from chains of ancilla atoms as indicated by red circles.
    (b) An MWIS problem is mapped to a UDG-MWIS and solved on the QPU, with the solution indicated by dashed grey circles.
    (c) This strategy can be extended to QUBO problems by transforming them into equivalent UDG-MWIS instances where weighted edges are embedded using quantum wires.}
  \label{fig:concept}
\end{figure}

Due to their ability to realize programmable spin models, neutral atom quantum computers have recently emerged as a promising platform for quantum optimization \cite{henriet2020,morgado2021}. In these systems, the Rydberg blockade mechanism naturally enforces the independence constraint for atoms spaced below a given radius, resulting in a native embedding of the MIS problem on a particular family of geometrical graphs called unit disk graphs (UDGs) \cite{pichler2018MIS,pichler2018complexity}. The use of QAA to solve the MIS problem on UDGs with King's connectivity using hundreds of atoms has been experimentally demonstrated recently \cite{ebadi22,kim2024}, with an indication towards a superlinear quantum speedup for hard graph instances. However, subsequent work \cite{andrist2023} showed that MIS instances on UDGs with King's connectivity can be optimally solved for up to thousands of nodes within minutes by classical approaches (custom or generic commercial solvers) on commodity hardware, without instance-specific fine-tuning. This result highlights the importance of extending the range of problems that can be embedded in neutral atom arrays beyond UDG-MIS, not only to widen the scope of possible real-world applications of this technology, but also to identify instances of quantum advantage. In this spirit, several proposals have been put forward to map non-UDG MIS \cite{kim2022,byun2022,dalyac2023}, QUBO \cite{byun2024QUBO}, higher-order unconstrained binary optimization (HUBO) \cite{byun2024HUBO}, satisfiability \cite{jeong2023SAT} and general combinatorial optimization \cite{lanthaler2024} problems onto the UDG-MIS problem using ancillary atoms and requiring only global control.

Another possibility to encode problems beyond UDG-MIS in neutral atom hardware is to use local control fields \cite{goswami2024}. This enables the realization of differential weights in the vertices represented by the atoms, permitting the embedding of MWIS problems defined on UDGs \cite{deOliveiraPritchard2025}. Importantly, UDG-MWIS problems can be further mapped into more general combinatorial optimization problems such as non-UDG MWIS, QUBO or integer factorization using an approach based on modular gadgets to copy and delocalize the information of the logical variables, and create arbitrary non-local connections \cite{nguyen2023,BombieriPichler2024}. Alternatively, an implementation based on the parity architecture \cite{lechner2015} allows conversion of QUBO and HUBO problems into UDG-MWIS \cite{lanthaler2023}. Both of these approaches require at worst an $\mathcal{O}(N^2)$ overhead for a problem with $N$ logical variables \cite{nguyen2023,park2024}, significantly reducing the ancillary qubit count from $\mathcal{O}(N^6)$ of previous proposals using only global addressing \cite{ebadi22}. However, while there are recent proposals to optimize the layout of the embedded graphs \cite{schuetz2024}, the overhead required for the known mappings into UDG-MWIS still poses a challenge for the embedding of large-scale combinatorial optimization problems onto near-term neutral atom hardware. 

In this paper, we introduce, and experimentally realize, a hardware-efficient scheme to map non-UDG MWIS and QUBO problems into a natively embeddable MWIS problem defined on a UDG. Our approach, schematically depicted in \figref{fig:concept}, is based on chains of atoms in the blockade regime, which we will call \emph{quantum wires}, to mediate interactions between physically separated vertices. For problems that can be formulated on graphs with quasi-UDG connectivity, and which require only a handful of long-range interactions, this allows one to significantly lower the overhead in ancillary atoms with respect to the general embedding schemes discussed above. By engineering the weights of the atoms that form the wires, we further ensure the preservation of the spectrum of the original problem.

This goes beyond the wire architecture presented in \cite{byun2022,kim2022}, which is designed to encode unweighted problems and only preserves the ground state, requiring post-selection of the anti-ferromagnetic state of the wires to identify valid solutions. Our wire constructions rely only on the blockade condition being fulfilled between neighbouring atoms, easing the experimental requirements with respect to proposals that require precise control of the relative strength of the interactions \cite{goswami2024,qiu2020}. 

The paper is organized as follows. In \secref{sec:background} we review the implementation of UDG-MWIS problems on neutral atom hardware. Section \ref{sec:quantumwires_MWIS} introduces wire constructions that allow for the encoding of MWIS problems on non-UDG geometries, while \secref{sec:quantumwires_QUBO} explains how these wires can be adapted to encode QUBO problems as UDG-MWIS. \secref{sec:hardware} discusses the hardware implementation and chosen annealing protocol~\cite{deOliveiraPritchard2025}, while \secref{sec:results} presents examples of non-UDG MWIS and QUBO problems embedded into UDG-MWIS using these wires, followed by experimental demonstrations of quantum annealing to validate the approach. We conclude in \secref{sec:conclusions} with a summary of our results, and an outlook for future work.

\section{Background}
\begin{figure*}[t!]
    \centering
    \includegraphics[width=\textwidth]{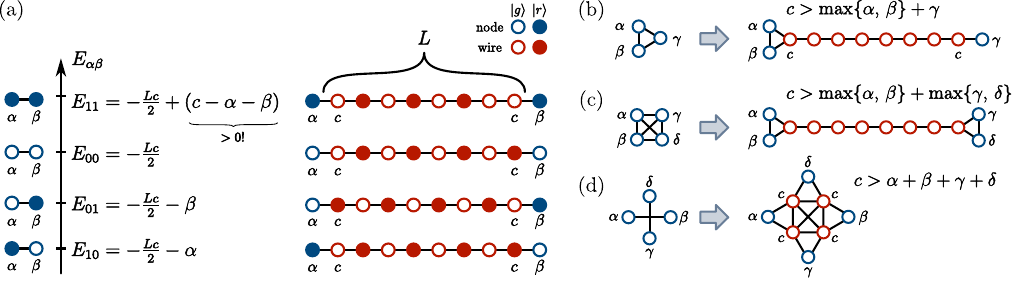}
    \caption[width=\textwidth]{{\bf Weighted Quantum Wires} (a) Basic construction of a wire to connect two nodes with weights $\alpha$ and $\beta$ in MWIS and QUBO problems. The energy diagram shows the ordering of the eigenstates for an MWIS implementation. (b, c) Wire constructions to delocalise triangular and all-to-all square interactions between vertices in MWIS problems. (d) Generalization of the crossing gadget introduced in \cite{nguyen2023} to allow for arbitrary weights of the nodes.}
    \label{fig:quantum_wire}
\end{figure*}
\label{sec:background}
We consider a weighted graph $G(V,E)$ which consists of a set of vertices $V$, each with weight $w_{i}>0$, and a collection of edges $E$ between pairs of vertices. The maximum weighted independent set is defined as the subset of vertices not connected by an edge that maximizes the total sum of the weights. Introducing binary variables $n_i=\{0,1\}$ to indicate whether vertex $i$ belongs to the set, we can identify the MWIS with the bitstring $\{n_i\}$ that minimizes the following classical cost function
\begin{equation}
H_{\text{MWIS}}=-\sum_i w_{i} n_i + \sum_{(i,j)\in E} U_{ij} n_i n_j ~,
\label{eq:costMWIS}
\end{equation}  
where $U_{ij} > \max\{w_i\}$ is a  penalty term that ensures that no two vertices connected by an edge simultaneously belong to the MWIS. Let us also consider a unit disk graph, which is a graph that can be embedded in a two-dimensional plane and such that there is an edge between a vertex pair $(i,j)$ if and only if their separation fulfills the condition $r_{ij} < R$, where $R$ is the so-called unit disk radius. The cost function of the UDG-MWIS problem, i.e. the MWIS problem defined on UDG graphs, can be naturally mapped to the many-body Hamiltonian of a neutral atom array with a laser driving between a ground state $\ket{g}\equiv \ket{0}$ and a Rydberg state $\ket{r}\equiv\ket{1}$
\begin{equation}
    H_{\text{Ryd}} / \hbar=\frac{\Omega}{2}\sum_{i}\sigma_i^x-\sum_{i}\Delta_i n_i + \sum_{i<j}V\left(r_{ij}\right)n_in_j ~,
    \label{eq:HamRyd}
\end{equation}
where $\Omega$ is the Rabi frequency, $\Delta_i$ is a site-dependent detuning from the $\ket{0}\leftrightarrow\ket{1}$ transition frequency, $n_i\equiv \ket{1}_i\bra{1}$ is an operator that counts the number of Rydberg excitations on site $i$, and $V(r)=C_6/r^6$ is the dipole-dipole van der Waals (vdW) interaction strength between two atoms separated by a distance $r$. Due to this interaction, each atom has a blockade radius $r_B^i=\sqrt[6]{C_6/\sqrt{\Omega^2+\Delta_i^2}}$ in which only one excitation can occur. In the limit $\Omega=0$, defining the smallest blockade radius as $r_B^{\text{min}}=\sqrt[6]{C_6/\max\{\Delta_i\}}$ and assuming $V(r>r_B^{\text{min}})\approx 0$, the ground state of $H_{\text{Ryd}}$ encodes the solution of a UDG-MWIS problem with weights $0\leq\tilde{w}_{i}=\Delta_{i}/\Delta_{0}\leq 1$, and unit disk radius $R \approx r_B^{\text{min}}$, realizing the constraint $V(r_{ij})>\max\{\Delta_k\}$ for all $(i,j)$ sharing an edge. Although the assumption $V(r>r_B)\approx 0$ is well justified due to the rapid decay of the dipole interactions with the atom separation, in practical implementations the interaction tails can have a non-negligible effect on the ground state of the system. Therefore, it is necessary to further impose the condition that the maximal unwanted interaction is less than the smallest weight in the problem, $\underset{(i,j)\notin E}{\text{max}} V(r_{ij}) < \underset{k}{\text{min}} ~ \Delta_{k}$. This guarantees that all pairs of vertices not connected by an edge are allowed to belong to the MWIS. 

The correspondence between $H_{\text{MWIS}}$ and $H_{\text{Ryd}}$ opens up the possibility to solve general optimization tasks with neutral atom quantum computers by first mapping them to UDG-MWIS instances~\cite{lanthaler2023, lanthaler2024, nguyen2023, byun2024QUBO, byun2024HUBO,park2024}, preparing the ground state via quantum annealing, and finally undoing the mapping to recover the solution to the original problem. 

\section{Quantum wires for MWIS problems}
\label{sec:quantumwires_MWIS}

\figref{fig:quantum_wire} (a) illustrates a basic wire construction to realize an MWIS constraint between two physically separated vertices with weights $\alpha$ and $\beta$ (with $\alpha>\beta$ in this example). The allowed configurations in the logical MWIS problem are $\ket{\alpha,\beta}=\{\ket{10},\ket{01},\ket{00}\}$, with corresponding energies $\{-\alpha,-\beta,0\}$. The wire (highlighted in red) consists of an even number of atoms $L$ with a uniform weight $c$, arranged in such a way that every atom is within the blockade radius of its two neighbors, forming an effective one-dimensional chain. As shown in the energy diagram in \figref{fig:quantum_wire} (a), choosing ancilla weights $c > \alpha + \beta$ yields the correct ordering of the states of the logical nodes in the three lowest-energy configurations of the system formed by the logical nodes and the wire, with the logical state $\ket{11}$ being energetically forbidden. Even though we only show one possible configuration for the $\ket{00}$ and $\ket{11}$ states, we note that, without considering the effect of vdW interactions, they have respective degeneracies of $L/2 + 1$ and $L/2$ due to the degree of freedom corresponding to the localization of the domain wall \cite{BombieriPichler2024}.
All the states of the wire are shifted by a constant offset $-Lc/2$ with respect to their corresponding logical configurations, ensuring the preservation of the MWIS solution when the wire is embedded in a larger graph with more atoms. We emphasize that this construction \emph{preserves} the spectrum of the logical problem, and truly facilitates an interaction between physically spatially separated nodes.

The construction that enables us to realize an edge between two spatially separated weighted atoms can also be generalized to delocalize other connectivity structures involving a larger number of atoms. In \figref{fig:quantum_wire} (b) we illustrate a wire construction that realizes an all-to-all connectivity between three atoms with weights $\alpha,\beta$ and $\gamma$, where atoms $\alpha$ and $\beta$ are within the blockade radius of each other but outside the one of atom $\gamma$. The wire is constructed in the same way as in \figref{fig:quantum_wire} (a), but now with vertices $\alpha$ and $\beta$ connected to one end of the wire, and node $\gamma$ to the opposite end. The analysis of the possible configurations of the wire shows that in order to preserve the correct ordering of the energy spectrum, one can choose a weight for the ancillary atom of $c>\max\{\alpha,\beta\}+\gamma$. Similarly, in \figref{fig:quantum_wire} (c) we show how to use a wire of even length $L$ to stretch an all-to-all connected square formed by four atoms with weights $\alpha, \beta, \gamma$ and $\delta$ into two separated regions, in one of which there is a direct edge between $\alpha$ and $\beta$ and in the other an edge between $\gamma$ and $\delta$. In this situation, a choice of the wire weight $c>\max\{\alpha,\beta\}+\max\{\gamma,\delta\}$ ensures the preservation of the logical spectrum. Analogously, this quantum wire method can be extended to delocalise clique structures of larger size than those depicted in Figs.~\ref{fig:concept}(b) and (c).

In order to implement more complicated structures with long-range connectivity, it might be necessary for quantum wires to cross. This can be achieved with a generalization of the crossing gadget introduced in Ref.~\cite{nguyen2023}, as shown in \figref{fig:quantum_wire} (d). In the most general case, in which the four vertices involved in the crossing have different weights $\alpha, \beta, \gamma, \delta$, choosing the weight of the crossing gadget ancillary nodes as $c> \alpha+\beta+\gamma+\delta$ guarantees the preservation of the spectrum of the logical problem. Further details and a discussion of the robustness of the proposed quantum wires can be found in Appendix~\ref{sec:robustness}.

While a systematic scheme to automate problem embeddings using quantum wires is outside of the scope of this paper, we can make a heuristic resource estimation to identify in which cases the wire approach could provide a more favourable ancilla scaling than the $\mathcal{O}(N^2)$ that the gadget and parity architecture incur into \cite{nguyen2023,lanthaler2023}. Assuming a graph with $N$ atoms arranged in a square lattice, a single wire would utilize in the worst case $\mathcal{O}(\sqrt{N})$ ancillary qubits to realize an interaction between distant atoms. Accounting for the fact that when $N_{\text{wires}}$ are present it might be necessary to introduce crossing gadgets that increase the ancillary atom count by up to $\mathcal{O}(N_{\text{wires}}(N_{\text{wires}}-1))$. A conservative estimate suggests that the wire embedding would be advantageous for graphs in which most of the edges admit a unit disk representation and $N_{\text{wires}}\sim\mathcal{O}(\sqrt{N})$ edges require long-range connections. This will serve as our operational definition of quasi‑UDGs. Quantum wires would also be particularly well-suited for problems composed of several UDG units that are connected between each other only through a very small number of nodes. Additionally, quantum wire constructions as presented in our work could be added to the toolkit for overhead reduction in the gadget embedding presented in \cite{schuetz2024}. 

\section{Quantum wires for QUBO problems}
\label{sec:quantumwires_QUBO}
Quadratic unconstrained binary optimization is a well-known NP-hard problem with a myriad of applications \cite{lucas2014ising,kochenberger14}. The solution of a QUBO problem can be formulated in terms of a bitstring $\{n_i\}$, with $n_i\in\{0,1\}$ that minimizes the classical cost function
\begin{equation}
H_{\text{QUBO}}=\sum_i h_i n_i+\sum_{i,j} J_{ij}n_in_j ~,
    \label{eq:cost_QUBO}
\end{equation}
where $h_i$ and $J_{ij}$ are coefficients that define the specific optimization problem to be solved. In the case when $h_i=-|h_i|<0 ~ \forall ~ i$ and $J_{ij}\ge0 ~ \forall ~ (i,j)$, $H_{\text{QUBO}}$ can be implemented within the quantum wire framework presented here. To understand this, let us consider the simplest case of a QUBO problem defined over two variables described by the cost function
\begin{equation}
    f(n_1, n_2) = -|h_1| n_1  -|h_2| n_2+J_{12} n_1 n_2 ~.
    \label{eq:QUBO_2bits}
\end{equation}
Comparing this expression with the wire spectrum shown in \figref{fig:quantum_wire} (a), we can see that by choosing $\alpha = |h_1|, ~ \beta = |h_2|$ and $c = J_{12}$, the states of the wire exactly mimic the QUBO cost function~\eqeqref{eq:QUBO_2bits} up to a constant energy offset $-Lc/2$ (for $L \geq 2$). Note that the constraint $c > \alpha + \beta$, necessary for the MWIS implementation, is now relaxed so the state $\ket{11}$ can appear in any order in the spectrum depending on the value of $J_{12}$. 
\\
Following this basic construction for a 2-bit QUBO, larger problems can be implemented by assigning an atom with relative weight $w_{i} = h_i$ to each logical variable $n_i$, and connecting the logical qubits with wires with homogeneous weights $c_{ij}=J_{ij}$. The building blocks, shown in Figs.~\ref{fig:quantum_wire} (b) and (c), could also be used for triangles or squares with uniform $J$ couplings, as depicted in \figref{fig:concept} (c). The crossing gadget of \figref{fig:quantum_wire} (d) could be used in the same way as for MWIS problems. While this approach for embedding QUBO problems is only scalable for instances with mostly local connectivity, in these scenarios it can provide an overhead reduction with respect to the proposals in Refs.~\cite{lanthaler2023, nguyen2023, byun2024QUBO}, which are designed for general graphs.

\section{Hardware implementation}
\label{sec:hardware}

\begin{figure}[t!]
  \includegraphics[width=\linewidth]{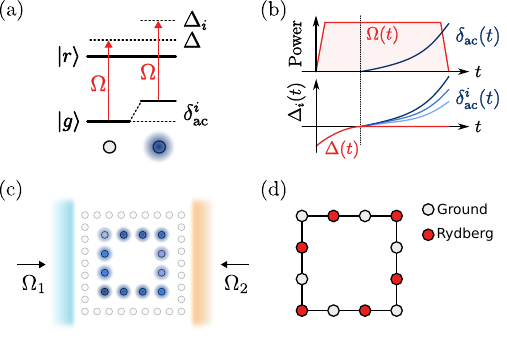} 
  \caption{{\bf Weighted annealing on an atomic QPU}. (a) local light-shifts encode vertex weights via $\delta_{ac}^i = w_i \delta_{ac}$. (b) The system undergoes annealing by globally ramping the detuning to resonance, followed by ramping of the local detunings to implement differential vertex weights. (c) Atoms are rearranged geometrically to encode MWIS vertices, with site-dependent light shift for weights. Rydberg excitation is performed using a pair of counter-propagating global excitation laser beams far detuned from the intermediate excited state ($\Delta/2\pi=1$~GHz) with an effective Rabi frequency $\Omega=\Omega_1\Omega_2/2\Delta$. Together this realizes the Hamiltonian described in Eq.~(\ref{eq:HamRyd}). (d) Post-annealing, Rydberg excitations are detected via atom loss.   }
  \label{fig:qpu_components}
\end{figure}

To implement MWIS annealing on neutral atom hardware, we employ the protocol introduced in \cite{deOliveiraPritchard2025}, using local light-shifts to implement node weights as illustrated in \figref{fig:qpu_components} and detailed in Appendix~\ref{appendix:experiment_setup}. This method integrates a global annealing schedule with site-specific, time-dependent light shifts, dynamically generated via an acousto-optic modulator (AOM) and a spatial light modulator (SLM). The protocol is defined by a global Rabi frequency $\Omega$ and local detuning pulses $\Delta_i(t)=\Delta(t)+\delta_{ac}^{i}(t)$, where $\Delta_i(t)$ is the global detuning of the Rydberg lasers and $\delta_\mr{ac}^i(t)=w_i\delta_\mr{ac}(t)$ encodes vertex weights through local light-shifts.

Atoms are spatially arranged to represent the unit-disk graph instance of the MWIS problem, with edge-connected qubits placed within a blockade radius ($r_B\sim\SI{12}{\micro\meter}$) to enforce interaction constraints. Initially, the system is prepared in the ground state $\ket{0}^{\otimes N}$ with $\Omega=0$ and $\Delta_i=\Delta<0$.
The Rabi frequency is increased to its maximum and kept constant while the global detuning $\Delta(t)$ is swept to resonance ($\Delta_i=\Delta=0$). After crossing resonance, the local detunings $\Delta_i(t)=\delta_{ac}^{i}(t)=w_{i}\delta_{ac}(t)$ are introduced for every atom while keeping the global Rydberg lasers at resonance ($\Delta=0$). During annealing, the maximum local detuning is limited to $\Delta_i^\text{max}(t)< 0.9 C_6/R^6$ to maintain blockade constraints between edge-connected atoms. Annealing typically occurs over a timescale of about \SI{8}{\micro\second}. This procedure is illustrated in \figref{fig:qpu_components} (a) and (b). The local detunings are calibrated via closed-loop feedback of the SLM hologram using spectroscopic measurements, typically achieving $\lesssim7\%$ RMS error after five iterations. The annealing schedule ensures that blockade constraints are satisfied as detunings increase, and the final Hamiltonian corresponds directly to the classical MWIS cost function of Eq.~\eqref{eq:costMWIS}. 

After annealing, Rydberg excitations are detected via atom loss. Repetition of the annealing sequence multiple times, combined with excluding trials that showed imperfect loading, generates statistics and configurations are ranked by occurrences. When the annealing protocol is executed properly, the most frequent configuration typically corresponds to the many-body ground state of the system. Errors arising from blockade violations, losses and imperfect state preparation and readout can be identified and corrected through classical post-processing, following the protocol introduced in \cite{ebadi22}. We first apply vertex reduction to eliminate vertices that violate the edge constraints of the graph to recover a valid independent set solution string. In many cases, this corrected configuration permits additional excitations without violating blockade constraints. The insertion of those excitations is performed by a fixed-depth vertex addition algorithm. See Appendix~\ref{appendix:post_processing} for further details. 
Closed-loop optimization can be used to tune the annealing parameters by seeking to minimize the weighted average cost $\langle H_\mathrm{MWIS}\rangle$ 
calculated over the ensemble of observed configurations.

\section{Results and Discussion}
\label{sec:results}

\subsection{Quasi-UDG MWIS}
\label{subsec:quasiudg_mwis}

\begin{figure}[t!]
  \includegraphics[width=\linewidth]{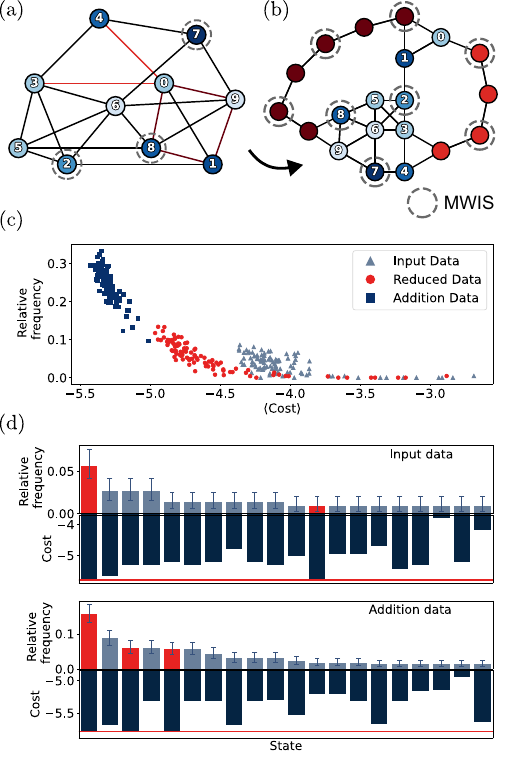} 
  \caption{Solving non-UDG MWIS problems using quantum wire encoding. 
  (a) A non-unit disk graph instance of the maximum weighted independent set problem is first rearranged, 
  and quantum wires are inserted to transform it into a UDG. 
  (b) The resulting UDG can be natively implemented on a neutral atom array.
  In this example, two weighted quantum wires are added to realize the connections highlighted in red. The MWIS solution is indicated by dashed lines around Rydberg excited nodes in the logical and embedded graphs.
  (c) Relative frequency of ground state observations versus average MWIS cost observed during annealing parameter optimization.
  Classical post-processing of the measured configurations enhances the anti-correlation between these quantities.
  (d) Histograms of the 20 most frequent states and their cost from 230 (out of 308) defect-free realizations of the annealing run yielding the lowest (vertex addition) average cost.
  Ground state configurations, which correspond to MWIS solutions, are highlighted with red bars, while their cost is highlighted by a red horizontal line.
  The top histograms shows raw experimental data, while the bottom displays frequencies and costs after applying both vertex reduction and addition.
  }
  \label{fig:scenario3}
\end{figure}

The first sparse graph example for which we want to find the MWIS solution is shown in \figref{fig:scenario3} (a) featuring 10 vertices. The set of vertex weights used in this problem instance is given by $\{w_0, w_1, \ldots, w_9\} = \{0.22, 0.51, 0.36, 0.21, 0.46, 0.22, 0.1, 0.57, 0.48, 0.1\}$.
By rearranging the positions of the nodes, all the edges can be made to satisfy the UDG criterion except for the triangle between nodes $\{0,4,5\}$, and the all-to-all connected cluster formed by nodes $\{0,1,8,9\}$. As shown in \figref{fig:scenario3} (b), these links can be realized with quantum wire constructions, resulting in an embedded graph formed by a total of 20 atoms with exclusively UDG connectivity. For comparison, we note that the implementation of the same logical problem using the gadget encoding described in \cite{nguyen2023} with additional optimization takes a total of $211$ atoms using the open software package \texttt{UnitDiskMapping.jl} \cite{UDmapping}. Naturally, this could be further optimized, for example using graph reduction techniques \cite{schuetz2024, SchuetzKatzgraber2025}, but for quasi-UDGs our technique provides a robust and efficient alternative that can be readily solved with weighted-graph annealing in our neutral atom hardware. 

The ground state of the encoded problem is degenerate, with three distinct configurations corresponding to the minimum MWIS cost. Among these, the configuration highlighted with dashed gray circles in \figref{fig:scenario3}(b) represents the ground state obtained using the experimentally calibrated interaction weights. To optimize the annealing schedule, we employed a closed-loop procedure that iteratively tuned the parameters of the annealing schedule. Classical post-processing of the measurement outcomes reveals a clear anti-correlation between the ground state relative frequency and the average MWIS cost, validating the effectiveness of the annealing parameter optimization as shown in \figref{fig:scenario3}(c). To further illustrate the performance of our approach, we highlight a specific measurement corresponding to the lowest average cost $\langle H_\mathrm{MWIS}\rangle$ observed during the optimization cycle. This measurement, shown in \figref{fig:scenario3}(d), confirms that the most frequent configuration is a valid ground state solution of the full MWIS graph. The histogram data is sorted first by frequency of occurrence, then by state index. We display only the first 20 bins for clarity, along with the cost of each measured state. In the data, we observe that whilst there should be three degenerate solutions, the system preferentially returns one of the configurations. This is likely due to small calibration errors in the relative weights ($<7\%$ RMS) and trap positions that break the degeneracy associated with different configurations of excitations along the quantum wire. However, since these all contain the same configuration of logical qubits, the broken degeneracy does not prevent observation of a valid solution to the target problem instance. 

Both the input graph and the post-processed data are shown, providing direct evidence of the system’s ability to identify MWIS solutions through quantum annealing on the neutral atom hardware. In the input data, we observe a target ground state solution with a relative frequency of $0.05\pm 0.02$. Without the annealing sequence (see Appendix~\ref{appendix:experiment_setup} for details), the expected probability of observing a target state solely due to atom loss is $p_\mathrm{target}=p^{(N-n)}(1-p)^n$, with target state length $N$, number of lost atoms $n$, and measured baseline survival rate $p$. Using $N=20$, $n=8$ and $p=0.985$, we find $p_\mathrm{target}=2\times10^{-15}$ for the MWIS solution. These results show that the coherent annealing process dominates over atom losses, and validates the approach of using weighted quantum wires for finding MWIS solutions on near term hardware.

\begin{figure}[t!]
  \includegraphics[width=\linewidth]{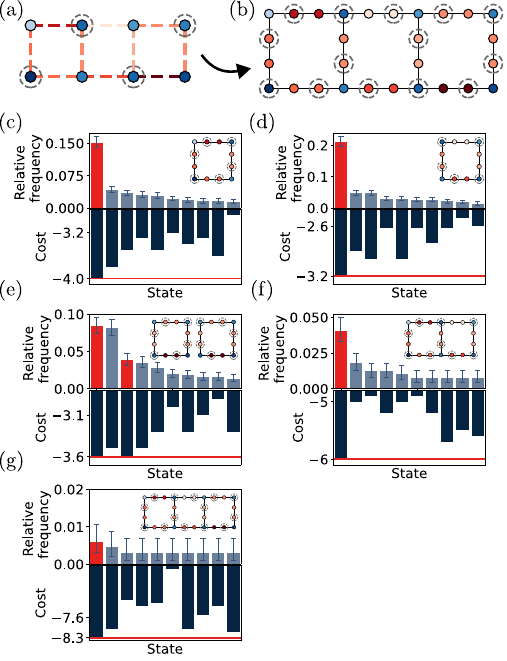} 
  \caption{Block-wise decomposition of a QUBO problem.
  (a) A QUBO problem consisting of 8 nodes and 10 edges is shown.
  This problem is mapped to a unit disk graph (UDG) formulation of the MWIS, 
  enabling native implementation on a Rydberg quantum processing unit.
  The QUBO graph can be partitioned into blocks of varying sizes by selectively removing edges. 
  (c–e) Annealing results for each individual block derived from (a), shown from left to right. 
  Block (e) exhibits two degenerate ground state solutions. Data shown for 870, 763 and 749 defect-free realizations respectively  out of 1000 trials.
  (f) MWIS results corresponding to the QUBO problem composed of the two leftmost blocks (664/1000 defect-free realizations). 
  (g) MWIS results for the full three-block QUBO problem for 669 defect-free realizations. In (c-g) the ground state configurations are highlighted in red.
  }
  \label{fig:qubo_blocks}
\end{figure}

\subsection{QUBO Problem Graphs}
\label{subsec:qubo_results}

To demonstrate that quantum wires can be used to natively embed QUBO problems on neutral atom arrays, we investigate small QUBO problems structured as square-shaped graphs, which act as modular components for constructing more complex instances. For each QUBO, the problems are generated using randomized node weightings and edge couplings. As illustrated in \figref{fig:qubo_blocks}(a), these blocks can be combined up to three at a time to generate larger problem configurations. Using a top-left to bottom-right indexing convention, the QUBO problem depicted in \figref{fig:qubo_blocks}(a) is characterized by the linear coefficients $\{h_0, h_1, \dots, h_7\} = \{-0.3, -0.8, -0.6, -0.7, -1.0, -0.7, -0.8, -0.9\}$ and couplings $\{J_{01}, J_{04}, J_{12}, J_{15}, J_{23}, J_{26}, J_{37}, J_{45}, J_{56}, J_{67}\} = \{0.8, 0.5, 0.1, 0.4, 0.4, 0.3, 0.4, 0.5, 0.6, 1.0\}$. Each QUBO instance is then translated into a UDG embedding using quantum wires as shown in \figref{fig:qubo_blocks}(b), ensuring operational compatibility with the neutral atom array. For each configuration we perform weighted-graph annealing and gather measurement statistics over multiple experimental runs, with the total number of defect-free realizations (those with all atoms present)  recorded in the caption.

As shown in \figref{fig:qubo_blocks} (c–g), the experimental outcomes consistently favor the ground state configurations which emerge as the most frequently observed results and correspond to the logical QUBO solutions. The histograms show the ten most common configurations  without any post-processing of the data for each graph and confirm that the MWIS solutions dominate the distributions. The post-processed data are presented in \figref{fig:qubo_blocks_addition} for reference.

\begin{figure}[htb!]
  \includegraphics[width=\linewidth]{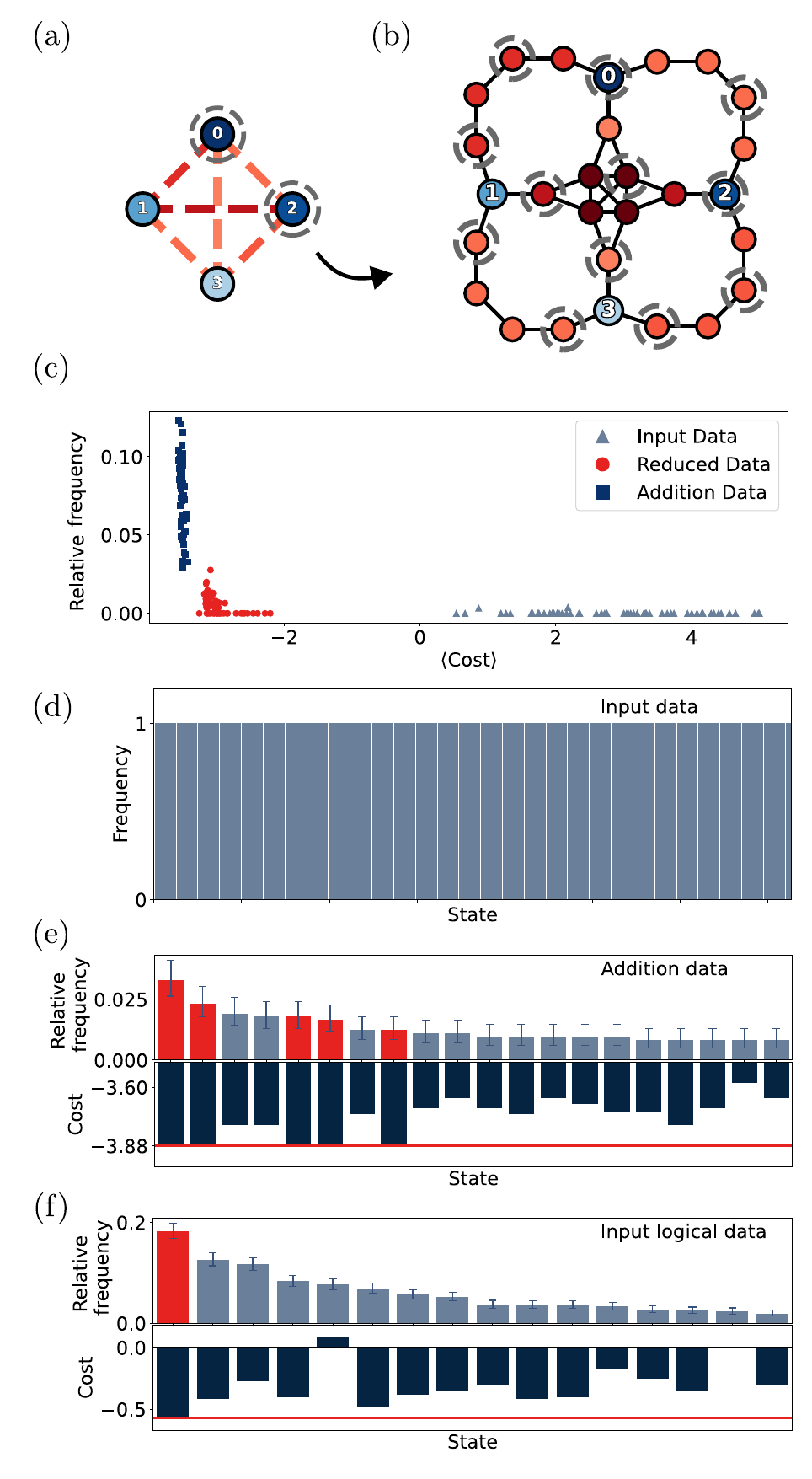} 
  \caption{Solving a fully connected QUBO problem via UDG-MWIS mapping. (a) A QUBO instance with 4 nodes and all-to-all connectivity is shown. (b) This problem is mapped to a UDG-MWIS using quantum wires and a modified crossing gadget. The resulting system exhibits six degenerate ground states with distinct wire configurations but identical QUBO solutions. (c) Ground state relative frequency versus average MWIS cost observed during annealing parameter optimization. (d) Histogram showing the raw frequency of all the observed states out of 728 (out of 2000) valid repetitions of a representative annealing run. (e) Histogram for the relative frequencies and costs of the 20 most observed states after applying both vertex reduction and addition to the data. The two most likely configurations are ground states (red) and can be mapped back to the solution of the all-to-all QUBO problem. (f) Histograms showing the raw frequency and cost of all 16 possible states for the four logical sites enumerated in (b). }
  \label{fig:qubo_ata}
\end{figure}

To further demonstrate the capabilities of our approach, we now consider a fully connected QUBO problem consisting of four weighted logical nodes $\{h_0, h_1, h_2, h_3\} = \{-0.4, -0.3, -0.375, -0.25\}$ linked by six weighted edges $\{J_{01}, J_{02}, J_{03}, J_{12}, J_{13}, J_{23}\} = \{0.275, 0.2, 0.175, 0.325, 0.2, 0.225\}$ as illustrated in \figref{fig:qubo_ata}(a). This densely connected graph is mapped onto a UDG-MWIS embedding using a combination of constructions from \figref{fig:quantum_wire} (a) and (d). The resulting layout requires 28 atoms: 4 for the logical vertices and 24 auxiliary atoms forming both standard quantum wires and a crossing gadget for overlapping connections. \figref{fig:qubo_ata}(b) shows the UDG-MWIS embedding used to enable graph optimization on the QPU.

We applied the closed-loop optimization protocol described earlier to refine the annealing ramp parameters, as shown in \figref{fig:qubo_ata}(c). Analysis of the measurement data through classical post-processing revealed a pronounced anti-correlation between the ground state relative frequency and the average cost throughout the optimization cycle. In contrast, the raw data shows consistently flat histograms with each experimental shot returning a single unique bitstring. Surprisingly, even for the data set giving the highest post-processed target state relative frequency (shown in \figref{fig:qubo_ata}(d) and (e)), we never observe any of the 6 possible target solutions for the full 28 qubit MWIS in the input data. Despite not returning a full ground state solution, tracing over the four logical qubits we measure the correct logical state with a relative frequency of $0.183 \pm 0.01$ from the raw input data (see \figref{fig:qubo_ata}(f)).

Whilst this example is far from a demonstration of quantum advantage, as a classical solver could readily solve the underlying 4-bit QUBO, it is instructive to quantify the performance of the neutral atom QPU in providing MWIS solution samples to the post-processing algorithm. We compare the QPU performance to an equivalent number of samples (728) chosen uniformly from the $2^{28}$ possible output strings as inputs to the classical post-processing algorithm. Averaging over multiple repeats, we find that the QPU provides ground state solutions with a relative frequency of $0.11\pm 0.01$, whilst the random sampling provides only $0.02\pm 0.005$, showing the QPU is providing higher quality samples despite experimental imperfections. As the time-to-solution scales as $-1/\log(1-p_\mathrm{solution})$ \cite{andrist2023}, the ability to post-process the raw data to obtain solutions with higher $p_\mathrm{solution}$ enables us to recover useful sample data even where the full state is never perfectly prepared. See Appendix~\ref{appendix:qpu_as_sampler} for more details.

These results reinforce the relevance of classical post-processing in extracting meaningful solutions from noisy quantum annealing outputs from the QPU at moderate system sizes, as have been previously used in other demonstrations of quantum annealing for atomic \cite{ebadi22} and superconducting hardware platforms \cite{koch25}.

\section{Conclusion}
\label{sec:conclusions}
In this paper, we proposed and experimentally demonstrated a novel quantum wire architecture to map certain hard combinatorial optimization problems on general graphs to a UDG-MWIS problem. This quantum wire encoding serves as a building block for solving optimization problems on neutral atom quantum processors, allowing for an efficient encoding and greater flexibility in problem graphs that can be solved using this technology. 

We demonstrated how our quantum wire scheme is able to efficiently encode MWIS problems beyond the UDG connectivity by mediating effective interactions between distant vertices. Furthermore, by carefully tuning the weights of the ancillary wire atoms, we are able to faithfully preserve, and tune, the entire logical spectrum of the target problem which allows for successful native encodings of QUBO problems within the same architecture. Because we do not encode a single qubit in the entire wire, but rather we use it to facilitate long-range connections between distant nodes, we find that, relative to previous general schemes \cite{nguyen2023, lanthaler2023}, our protocol requires substantially fewer resources.

Beyond the theoretical framework, we experimentally implement our method by solving non-UDG maximum-weight independent set and QUBO problems via quantum annealing on neutral atom arrays. These results confirm the feasibility of our embedding strategy and its compatibility with current neutral atom hardware.

Finally, we note that while we discuss this embedding scheme from a physical implementation perspective, it can also be used as a sub-routine for classical solvers to efficiently map non-unit disk graphs into weighted UDG problem instances. An intriguing direction for future work is the optimization and automatization of the embedding layout, as well as the potential integration of the wire architecture with existing toolkits for the reduction of MWIS problems and the identification of hard instances that would benefit from the implementation of quantum solutions \cite{schuetz2024,SchuetzKatzgraber2025}.

\begin{acknowledgements}
We thank Gillian Marshall and Kristina Bell from Qinetiq for useful discussions and provision of some of the example problems that underpin the results presented above. We also thank C. J. Picken for his technical support. This work was supported by EPSRC through the Prosperity Partnership \emph{SQuAre} with funding from M Squared Lasers Ltd. (EP/T005386/1) and grants EP/Y005058/2 and EP/Z53318X/1, as well as Innovate UK through the Innovation Accelerator for Neutral Atom Quantum Optimization (Grant No. 10059444). The data presented in this work are available at \cite{pureDOI}. 
\end{acknowledgements}

\appendix
\section{Robustness of the wire architecture}
\label{sec:robustness}
While the proposed wire architecture is guaranteed to give the correct ground state solution, it is \emph{a priori} not clear how robust the annealing ramps and the embedding scheme are to experimental imperfections and uncertainties. In this Appendix we separately consider two aspects of this - $(1)$ how robust and scalable the annealing performance of the quantum wires are as the system sizes is increased, and $(2)$ how resilient the embedding is to variations in the vertex weights.

\subsection{The annealing protocol}
As indicated by the results discussed above, the quantum wire approach presented here allows for a flexible and robust encoding of combinatorial optimization problems that are close to UDG connectivity. In order to better understand the reason for this robustness, we numerically compute the spectral gap of the basic building blocks in this architecture as a function of system size.

\begin{figure*}[t!]
    \includegraphics[width=\textwidth]{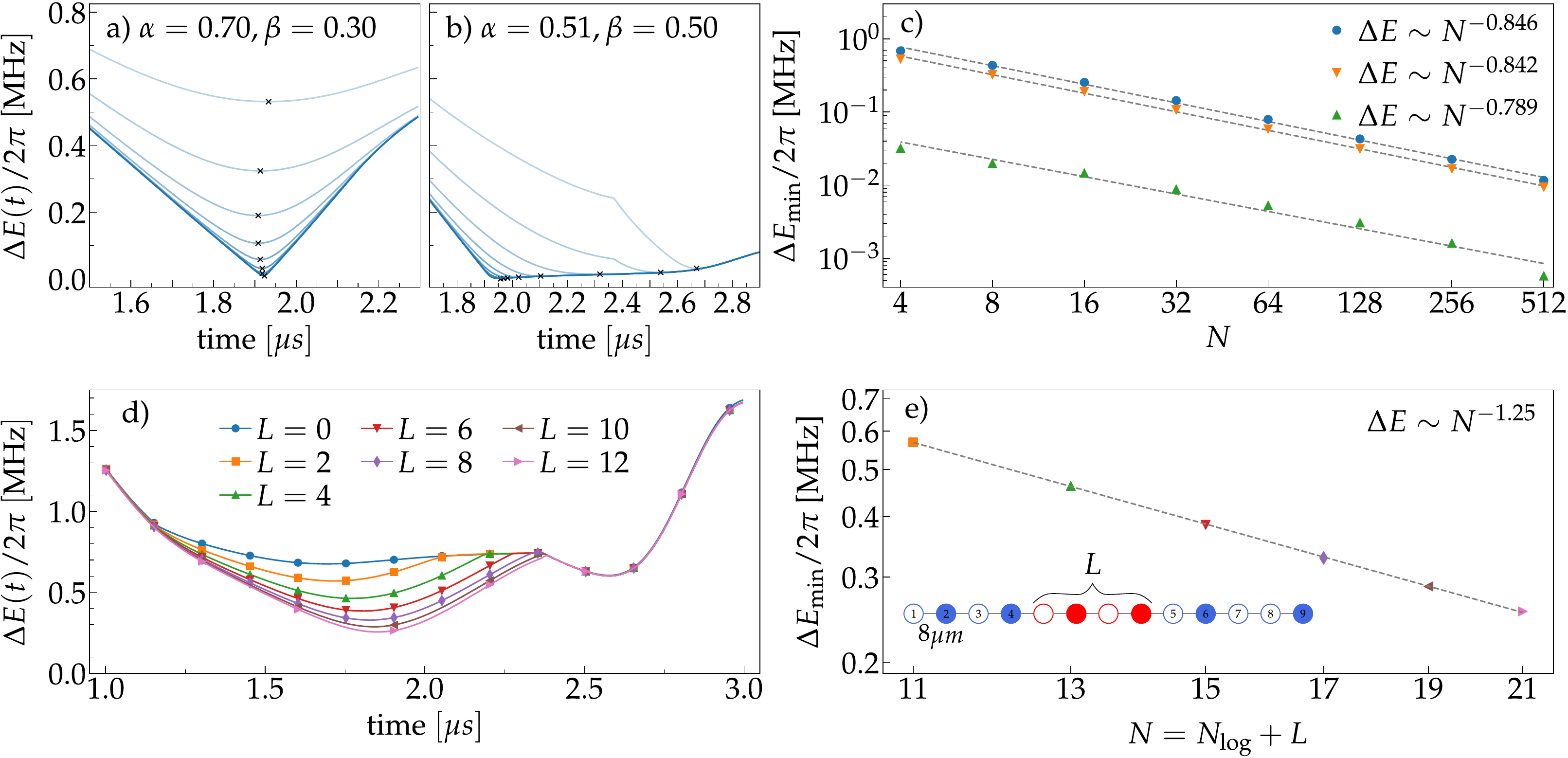}
    \caption{The left two plots show the spectral gap along the annealing path $\Delta E(t)$ for different system sizes $N = \{4, \ldots,512\}$ (light to dark blue) for wire weights $(\alpha,\beta) = (0.7,0.3)$ (a) and $(\alpha,\beta) = (0.51,0.5)$ (b). The black markers indicate the extracted minimum spectral gap depicted in the main plot. (c) Minimal spectral gap $\Delta E_{\mathrm{min}}$ as a function of system size for wire weightings $(\alpha,\beta) = \{(0.1, 0.9), (0.7, 0.3), (0.51, 0.5)\}$ (blue circle, orange inverted triangle, and green triangle respectively). The spectral gap closes polynomially with system size $\sim N^{-z}$ , with a linear fit giving an exponent of $z \lesssim 1$. These results were obtained using DMRG simulations using a bond dimension of $D=200$, and a truncation error of $\epsilon = 10^{-12}$. The extracted energies were converged to a threshold of $10^{-4}$. (d) Spectral gap along annealing path for the wire in (a) embedded in a 1D weighted logical graph of $N_{\textrm{log}} = 9$ nodes for varying lengths $L$ of the wire (note $L=0$ shows the gap of just the logical graph). For clarity only a subset of the data points are indicated by the markers. (e) Minimal gap along the annealing path for different system sizes extracted from (d). Here too, the minimal spectral gap of a logical problem embedded using a quantum wire closes polynomially as $\sim N^{-z}$ as the wire length is increased. The full system (logical graph and quantum wire) is depicted in the inset for a wire length of $L=4$. These results were obtained using exact diagonalization, and a linear fit to the minimal spectral gap yields an exponent of $z \sim 1.25$.}
    \label{fig:quantum_wire_gap}
\end{figure*}

The two leftmost panels of \figref{fig:quantum_wire_gap} show the spectral gap $\Delta  E(t)$ of the Rydberg Hamiltonian Eq.~\ref{eq:HamRyd} along the annealing path for logical wire weightings $(\alpha,\beta) = \{(0.7,0.3), (0.51, 0.5)\}$ in (a) and (b) respectively. The different shadings indicate the system size (logical and ancilla atoms) of the wire $N = 2^{k}$ for $2 \leq k \leq 9$. The minimal spectral gap $\Delta E_{\mathrm{min}}$ along the path is marked by black crosses for each system size. We used the same annealing schedule for all considered logical weights. It has been variationally optimized for the smallest $N=4$ wire with logical weights $(0.7, 0.3)$ to isolate the effect of the wire length on the spectral gap. Using the matrix product state (MPS) ansatz, we variationally find the spectral gap using the density matrix renormalization
group (DMRG) algorithm \cite{White1992, White1993, Schollwoeck2011, ITensor2022,ITensor-r0.3}. 
\figref{fig:quantum_wire_gap} (c) shows the minimum spectral gap (as indicated by the black markers in the left plots) as a function of system size $N$ for different wire weights. For the wires, we find that the spectral gap closes polynomially as $N^{-z}$ with $z \lesssim 1$, indicating that these wires do not pose a fundamental problem for the adiabatic protocol. These findings are also in agreement with the recent analysis performed in Ref. \cite{BombieriPichler2024}. 

The black markers in \figref{fig:quantum_wire_gap} (a) and (b) indicate where the minimal spectral gap occurs along the annealing path. We have chosen two distinct scenarios of a strongly imbalanced (a) and a closely balanced (b) wire. We already saw above that the spectral gap closes more slowly (smaller $z$) for the former case (a). Interestingly, in this case the minimum gap appears to occur around the same detuning value, independent of the wire length. Meanwhile for the closely balanced wire (b) the gap closes more quickly with system size, and there is a larger variation in the place where the minimal gap occurs along the annealing path. This is intuitively expected as the MWIS problem is computationally more difficult if the weights are all very similar due to smaller energy gaps between the different configurations. We obtain qualitatively very similar results for the triangle and square constructions of \figref{fig:quantum_wire} (b-c) with a scaling exponent of $z \lesssim 1.5$.

Finally in order to study how these quantum wires perform in a real logical graph, we choose a simple one-dimensional MWIS problem of $N_{\textrm{log}} = 9$ nodes separated by a lattice spacing of $d = \SI{8}{\micro\meter}$. The logical weights of the problem are given by the vector $\vec{w} = \{0.22, 0.87, 0.21, 0.7 , 0.3 , 0.92, 0.49, 0.61, 0.77\}$, and we used the quantum wire of \figref{fig:quantum_wire_gap} (a) to connect the leftmost $4$ to the rightmost $5$ logical nodes. \figref{fig:quantum_wire_gap} (d) shows the spectral gap along the annealing path (same annealing schedule as used in the previous simulations) for varying lengths of the quantum wire $L = 0, \ldots 12$. Extracting the minimal spectral gap as above, we can see in (e) that as we make the wire longer, the spectral gap of the full system still closes polynomially with system size $N = N_{\textrm{log}} + L$, with a scaling exponent of $z \sim 1.25$. We are therefore encouraged that the quantum wires will not lead to adiabatic bottlenecks when annealing real graphs, due to the benign closing of their spectral gap.

\subsection{Robustness of a single quantum wire}
In order to evaluate how robust the embedding scheme is to shot-to-shot fluctuations in the applied local light-shifts, we isolate a single wire module, and vary the local detunings randomly. For concreteness we draw the random weights from a normal distribution centred around the ideal detuning value, and study how these deviations from the ideal implementation can affect the low-energy spectrum of the quantum wire. Let $X$ be a random variable $X \sim \mathcal{N}\lr{\mu_{X}, \sigma_{X}^{2}}$, where $\mu_{X} = \{\alpha, \beta, c \}$, and $\sigma_{X} = \{\sigma_{\alpha}, \sigma_{\beta}, \sigma_{c} \}$ respectively. The configurational energies of the wire are thus also random variables, and specifically the distributions of the low-energy states shown in \figref{fig:quantum_wire} (a) are given by $E_{x_{\alpha},x_{\beta}} \sim \mathcal{N}\lr{\mu_{x_{\alpha},x_{\beta}},\sigma^{2}_{x_{\alpha},x_{\beta}}}$, with 

\begin{eqnarray}
    \mu_{x_{\alpha},x_{\beta}} &=& - \alpha x_{\alpha} - \beta x_{\beta} - c N_{x_{\alpha},x_{\beta}} ~, 
    \label{eq:mean_energies_random}\\
    \sigma^{2}_{x_{\alpha},x_{\beta}} &=& x_{\alpha}\sigma_{\alpha}^{2} + x_{\beta}\sigma_{\beta}^{2} + N_{x_{\alpha},x_{\beta}} \sigma_{c}^{2} ~,
    \label{eq:stdv_energies_random}
\end{eqnarray}
their mean and variance respectively. Here $x_{i} \in \{0,1\}$ indicates the absence (presence) of an excitation on either end of the wire ($\alpha$ or $\beta$ respectively), and $N_{x_{\alpha},x_{\beta}} = (L - 2x_{\alpha}x_{\beta})/2$ is the total number of excitations on the ancilla atoms for the considered states.
The expectation values of the energies coincide with the ideal wire result as expected, and we note that fluctuations in the detunings lead to a standard deviation $\sigma = \sqrt{\sigma_{\alpha}^{2} + \sigma_{\beta}^{2} + L \sigma_{c}^{2}}$ which scales as $\sigma\sim\sqrt{L}$ with increasing wire length. We also note that this analysis holds both considering only UDG interactions and van der Waals (vdW) tails, as in the latter case the long-range interactions are identical in the lowest two energy configurations.

\begin{figure}[t!]
    \includegraphics[width=0.5\textwidth]{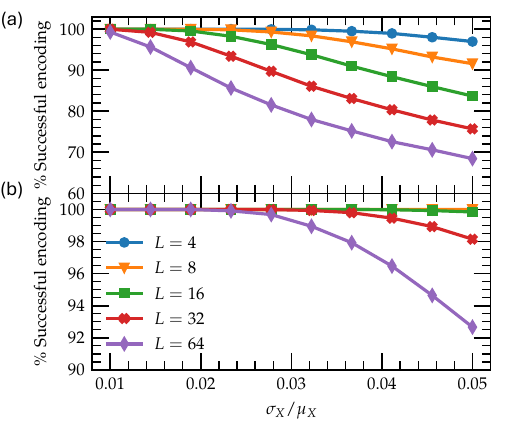}
    \centering
    \caption{Probability of successfully encoding the ground state in the presence of Gaussian fluctuations in the local light-shift for wires of different length $L$ as a function of the standard deviation $\sigma_X$ of the distribution. Plot (a) corresponds to the choice of weights $\alpha=0.6, \beta=0.4$, and (b) $\alpha=0.8,\beta=0.2$. Each data point has been estimated by sampling across 100,000 random realizations of the local light-shifts.}
    \label{fig:histogram_randomweights_wire}
\end{figure}

We test empirically the robustness of a single wire to noise in the local light-shifts by setting the ground state with $\alpha>\beta$, randomizing the weights of the logical nodes and the wire atoms in the manner discussed above and checking whether the new ground state still corresponds to the $\ket{10}$ state. To speed up the calculations, we only evaluate the energies of the $L+3$ configurations of the system formed by the logical nodes and the wire that respect the blockade constraint. In Fig. \ref{fig:histogram_randomweights_wire} (a) and (b) we show respectively the results corresponding to $\alpha=0.6, \beta=0.4$ and $\alpha=0.8, \beta=0.2$ for wires of different lengths $L$ and for different values of the standard deviation $\sigma_X$ of the noise Gaussian, which for each atom is chosen such that $\sigma_X/\mu_X$ is constant across the system. Each data point has been obtained by sampling 100,000 random local detuning realizations. The success probability is larger for the case $\alpha=0.8, \beta=0.2$ shown in Fig. \ref{fig:histogram_randomweights_wire} (b) than for the case $\alpha=0.6, \beta=0.4$ of Fig. \ref{fig:histogram_randomweights_wire} (a). This is to be expected from Eq. \eqref{eq:mean_energies_random}, as the expectation value of the gap between the $\ket{10}$ and the rest of states is larger for $\alpha=0.8, \beta=0.2$ than for $\alpha=0.6, \beta=0.4$. Even though there is a decay in the probability of successfully encoding the solution of the MWIS problem with increasing values of the standard deviation $\sigma_X$ and the wire length $L$, for lengths of up to $L=64$ and standard deviation $\sigma_X/\mu_X=0.05$ the success probability remains lower-bounded by $\sim 70\%$ in the case $\alpha=0.6,\beta=0.4$ and by $\sim 92\%$ for $\alpha=0.8,\beta=0.2$. When the wire is embedded in a larger graph, we expect the robustness of the encoding to be enhanced by the additional gaps induced by the structure of the logical problem and the presence of residual long-range vdW interactions.

\subsection{Robustness of the crossing gadget}
To complete the analysis of the robustness of the different building blocks involved in our architecture, we also test the performance of the crossing gadget in the presence of random shot-to-shot fluctuations in the values of the weights and compare it with that of the quantum wires. As before, we sample the weights $\{\alpha,\beta,\gamma,\delta,c\}$ of the crossing gadget shown in Fig. \ref{fig:quantum_wire} (d) from normal distributions $\mathcal{N}(\mu_X,\sigma_X^2)$ such that the ratio $\sigma_X/\mu_X$ is constant across the system, and we study how likely is the crossing gadget to encode the correct ground state as a function of the standard deviation $\sigma_X$. In order to make a fair comparison with the robustness of the wires, we have picked different examples of logical weight configurations and compared with wires of length $L=2$ and $L=4$ with logical weights such that the gap of the logical states of the wires is the same as for the states of the gadget. The reason to choose these wire lengths is that the crossing gadget has four ancillary atoms but only one of them is excited in the ground state, so it makes sense to compare the robustness both with a wire with the same number of ancillary atoms ($L=4$) and the same number of ancillary excitations ($L=2$).

\begin{figure}[t!]
    \includegraphics[width=0.5\textwidth]{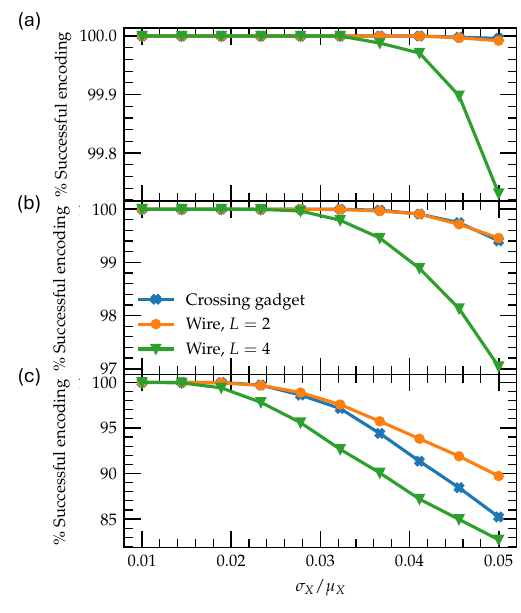}
    \caption{Probability of successfully encoding the ground state in the presence of Gaussian fluctuations in the local light-shift for a crossing gadget and wires of length $L=2,4$ as a function of the standard deviation $\sigma_X$ of the distribution. Plot (a) corresponds to the choice of weights $\{\alpha=\gamma=0.4, \beta=\delta=0.1\}$ for the gadget and $\{\alpha=0.65,\beta=0.35\}$ for the wires, (b) corresponds to the choice of weights $\{\alpha=0.4,\gamma=0.3, \beta=0.2,\delta=0.1\}$ for the gadget and $\{\alpha=0.6,\beta=0.4\}$ for the wires, and (c) corresponds to the choice of weights $\{\alpha=\gamma=0.3, \beta=\delta=0.2\}$ for the gadget and $\{\alpha=0.55,\beta=0.45\}$ for the wires. Each data point has been estimated by sampling across 100,000 random realizations of the weights.}
    \label{fig:noise_crossgadget}
\end{figure}

The results of this analysis are summarized in Fig. \ref{fig:noise_crossgadget}. We have chosen different values of the weights intervening in the crossing gadget such that in all cases $c=\alpha+\beta+\gamma+\delta=1$, and that the gap between the logical ground and first excited states takes the values $\Delta E=0.3$ (Fig. \ref{fig:noise_crossgadget} (a)), $\Delta E=0.2$ (Fig. \ref{fig:noise_crossgadget} (b)) and $\Delta E=0.1$ (Fig. \ref{fig:noise_crossgadget} (c)). We have checked independently that the robustness of the gadget depends only on the logical gap and not the particular choice of the weights. For each of these crossing gadget weight instances, we have chosen values of the wire logical weights such that $c=\alpha+\beta=1$ and that the gap matches that of the logical states of the crossing gadget. For energy gaps $\Delta E=0.3, \Delta E=0.2$, we observe that the crossing gadget is as robust to fluctuations in the weights as a wire of length $L=2$. For $\Delta E=0.1$ the robustness of the crossing gadgets deviates slightly from that of the $L=2$ wire for higher values of the variance, but it still remains higher than the robustness of wire of length $L=4$ with the same spectral gap. Overall, this analysis allows us to conclude that the crossing gadget has a robustness to random shot-to-shot fluctuations in the values of the detunings comparable to that of wires of length at most $L=4$.

\section{Experimental setup}
\label{appendix:experiment_setup}
\begin{figure}[t!]
  \centering
\includegraphics[width=\linewidth]{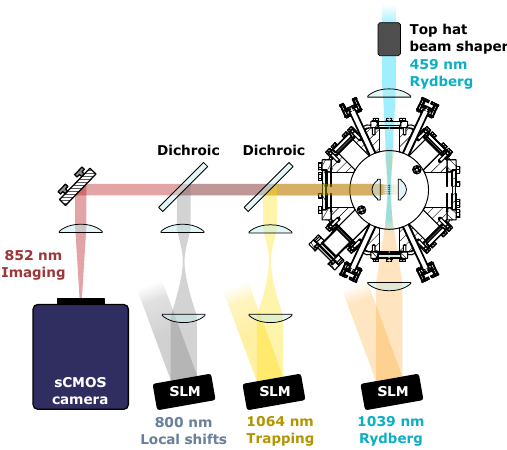} 
  \caption{Overview of experimental setup. The tweezer trap array is formed using an SLM to generate holographic tweezers using 1064~nm light, with detection performed using an sCMOS camera. Rydberg excitaiton is performed using global beams at 459~nm and 1039~nm, with uniform intensity created using top-hat beamshaper and an SLM respectively. local light-shifts are generated using an 800~nm beam imaged off an SLM to enable site-dependent weights.}
  \label{fig:setup}
\end{figure}

Our experimental system is shown schematically in \figref{fig:setup} and described in detail in \cite{deOliveiraPritchard2025}. An array of cesium (Cs) atoms are confined in holographically generated optical tweezers (1064 nm). To implement MWIS annealing, we combine global annealing protocols with the ability to implement arbitrary local light-shifts using an auxiliary spatial light modulator. Once the problem is mapped to UDG-MWIS, atoms are arranged spatially to represent the graph problem, where edge-connected qubits are placed within a blockade radius
of each other to ensure the strong interactions prevent simultaneous excitation of both qubits.

Atoms are initially prepared in the hyperfine ground state $\vert g\rangle=\vert 4,0\rangle$ and coherently transferred to the Rydberg state $\ket{r}=\ket{80s_{1/2},m_j=+1/2}$ using a two-photon excitation scheme via the $7p_{1/2}$ intermediate state.
The excitation is driven by counter-propagating laser beams at 459~nm and 1039~nm, each shaped into flattop intensity profiles to enhance uniformity across the atomic array. The 459~nm beam is flattened using a commercial beam-shaping optic 
(Osela DTH-1D-250µm-40µm-3.75mm), while the 1039~nm beam is shaped using an SLM \cite{schroff23}.

The excitation lasers are detuned from the intermediate state by $\Delta_\text{int}/2\pi=$1~GHz, enabling Rydberg coupling with a 
measured two-photon Rabi frequency of $\Omega/2\pi=1.0(5)$~MHz. Fine control of the global detuning is achieved through frequency modulation of 
the 459~nm beam using an acousto-optic modulator (AOM) in a double-pass configuration. For the $80s_{1/2}$ Rydberg level, the van der Waals
interaction strength is characterized by a coefficient $C_6=\SI{-3376}{\giga\hertz\micro\meter}^6$ \cite{sibalic17arc} which corresponds to a resonant blockade 
radius of $r_B=\sqrt[6]{\vert C_6\vert/\Omega}=\SI{12.1}{\micro\meter}$. To suppress atom loss and decoherence, 
the optical tweezers are turned off throughout the annealing sequence.

An additional light-shift potential introduces a site-dependent offset given by $\delta_\mr{AC}^i=w_i\delta_\mr{AC}$, where $w_i$ denotes the relative weight assigned to site $i$. The global term $\delta_\mr{AC}$ is governed by the total intensity of the light-shift laser, which is dynamically modulated using an AOM before reaching the SLM. This configuration enables the implementation of continuous annealing ramps while maintaining fixed relative weights throughout the experiment. These AC Stark shifts are realized using 800~nm light from a Ti:Sapph laser. Arrays of focused light-shift spots are projected onto atoms with a measured $1/e^2$ waist of $\SI{3.0(2)}{\micro\meter}$. 

\begin{figure}[t!]
  \includegraphics[width=\linewidth]{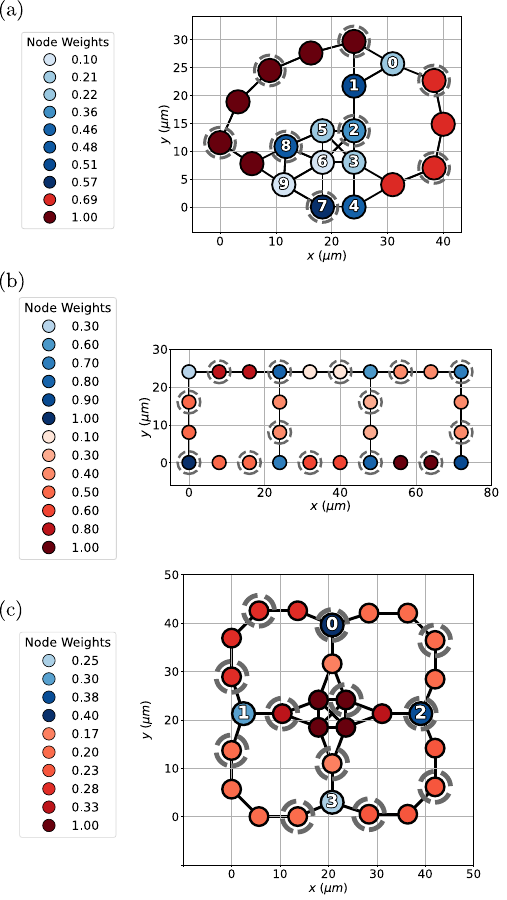} 
  \caption{Weighted graphs. (a) Graph of \figref{fig:scenario3}(b). (b) QUBO problem with weighted nodes and connections of \figref{fig:qubo_blocks}(a).
  (c) QUBO with all-to-all connectivity of \figref{fig:qubo_ata}(b).
  }
  \label{fig:weighted_graphs}
\end{figure}

\begin{figure*}[t!]
\includegraphics[width=0.75\linewidth]{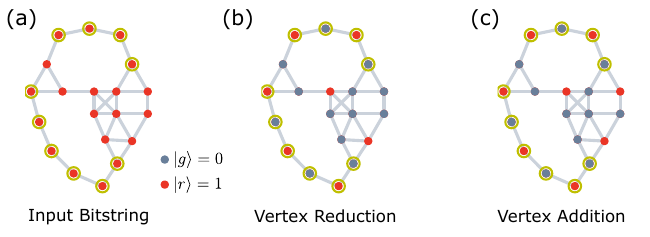}
\caption{{\bf Post-Processing Algorithms} (a) Input bitstring data containing edge (blockade) violations (b) Vertex reduction algorithm removes edge violations and recovers a valid independent set (IS) bitstring. (c) The vertex reduced bitstring may be an IS but have the option of adding additional vertices. Vertex addition uses a constant overhead greedy algorithm to randomly select vertices to convert from 0 to 1 if no edge violation occurs, and then returns the lowest cost solutions.}
\label{fig:classical}
\end{figure*}

For each instance of the UDG-MWIS problem, the intensity-weighted light-shift patterns are generated using a modified weighted-Gerchberg-Saxton 
phase retrieval algorithm, allowing precise control over the relative amplitudes of the individual sites \cite{kim2019large}. 
Calibration of the resulting Stark shifts is performed via spectroscopy on the $\ket{50s_{1/2},m_j=+1/2}$ Rydberg state, 
which is chosen to suppress interatomic interaction effects due to its lower principal quantum number. Measured shifts are used in a closed-loop feedback routine to iteratively update the SLM phase pattern until the RMS error in the target intensity distribution converges to a value below a certain threshold, typically $<7\%$.

To quantify the effect of atom losses, we perform a series of baseline survival measurements. The first measurement is to run the system with no 800~nm or Rydberg lasers active, from which we find a single atom survival of 99(1)\%. When adding the 800~nm light and running annealing ramps with no Rydberg light, this reduces to $p=98.5(1)\%$. Repeating the ramp with both 800~nm light and either the 459~nm or 1039~nm beams on, we see no change in atom loss, recovering $p=98.5(2)\%$ averaged over graph arrays with up to $N=28$ atoms.

\section{Graph weights}
\label{appendix:graph_weights}

\figref{fig:weighted_graphs} presents the weights used for all the UDG-MWIS problems investigated in the main text.

\section{Classical post-processing}
\label{appendix:post_processing}
These experiments have identified challenges in scaling up to larger problem sizes, namely minimizing effects of ground state loss which act as a false-positive for Rydberg excitation (and hence a higher chance of identifying configurations as high-cost blockade violations) and verification of interaction strengths as a function of array position to ensure the array spacings of the encoded problems are within the blockade radius. 

In the context of a many-body quantum simulation experiment these detrimental effects are significant, however in the context of using the atom array as an annealer for classical optimization it remains possible to take output bitstrings featuring edge-violations, and perform classical post-processing to recover valid graph solutions.  To explore the impact of this, we implement the same post-processing algorithms introduced by \cite{ebadi22} in the context of solving MIS problems, and apply these retroactively to our MWIS datasets.

The classical post-processing step acts to improve the quality of the final solutions with respect to the raw measured strings. It provides a simple mechanism by which to ensure firstly all strings become valid independent sets (removing edge-violations caused by neighbouring Rydberg states through vertex reduction), and secondly by maximizing the allowed number of excitations (hence lowering the cost) through vertex addition. The vertex reduction step can resolve errors caused by failures of adiabaticity (meaning preparation of higher energy states which violate the independent set (IS) condition) and atom losses, whilst the the vertex addition step can reduce errors due to imperfect state preparation or imperfect Rydberg detection that result in higher ground state detection than expected.

\subsection{Vertex reduction}
For experimental bitstrings featuring edge (or equivalently blockade) violations, whereby pairs of edge-connected vertices are both 
selected, we can recover valid independent set bitstrings by applying \textit{vertex reduction}. This algorithm is implemented 
by first counting the number of edge-violations on each vertex, and selecting the vertex with the highest number to 
be flipped from $1$ to $0$. Where many vertices all have the maximum number of edge violations, 
the vertex to be flipped is chosen at random. This process is repeated until all edge-violations are removed ensuring the
bitstring represents an IS. An example is shown in Fig.~\ref{fig:classical} where starting with the worst possible bitstring in (a) 
(all vertices selected), the edge violations are sequentially removed to obtain a valid IS bitstring shown in (b).

\subsection{Vertex addition}
Following the application of vertex reduction to the MWIS bitstrings we recover a valid IS but it may not be maximal, 
meaning it may be possible to select one or more additional vertices to recover a lower cost solution. 
A \textit{vertex addition} algorithm is implemented using a constant overhead greedy algorithm, whereby for a 
fixed number of iterations (in this case 10 as used in \cite{ebadi22}) the vertices are selected in a random order, 
and if a vertex is in state $0$ it is transformed to $1$ if it would not cause an edge violation. The lowest cost solution 
is then returned. An example is shown in Fig.~\ref{fig:classical} (c) where the algorithm has added an additional vertex to
achieve a lower cost.

\section{Vertex addition data for the block-wise QUBO problems}
\label{appendix:qubo_blocks_addition}
\begin{figure}[h!]
  \includegraphics[width=\linewidth]{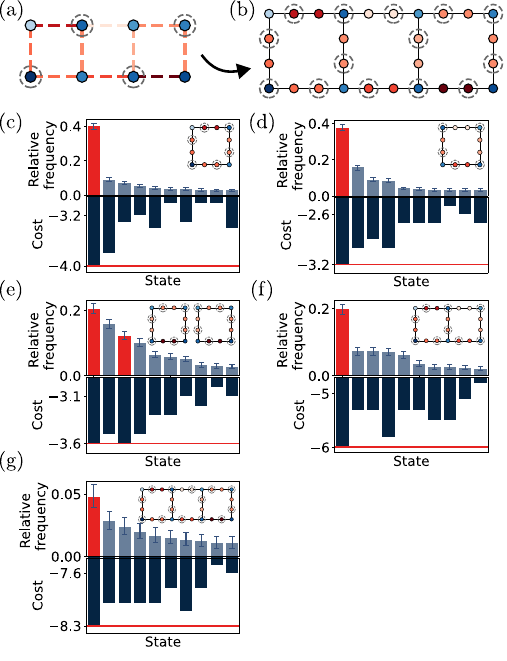} 
  \caption{Block-wise decomposition of a QUBO problem with post-processing. (a) A QUBO problem consisting of 8 nodes and 10 edges is shown. This problem is mapped to a UDG formulation of the MWIS, enabling native implementation on a Rydberg quantum processing unit. The QUBO graph can be partitioned into blocks of varying sizes by selectively removing edges. (c–e) Annealing results after vertex reduction and addition for each individual block derived from (a), shown from left to right. Block (e) exhibits two degenerate ground state solutions. Data shown for 870, 763 and 749 defect-free realizations respectively out of 1000 trials. (f) MWIS results corresponding to the QUBO problem composed of the two leftmost blocks (664/1000 defect-free realizations). (g) MWIS results for the full three-block QUBO problem for 669 defect-free realizations. In (c-g) the ground-state configurations are highlighted in red.}
  \label{fig:qubo_blocks_addition}
\end{figure}

Fig.~\ref{fig:qubo_blocks_addition} (c)-(g) presents the histograms obtained after post‑processing the data from Fig.~\ref{fig:qubo_blocks} (c)-(g), illustrating how classical post‑processing improves the relative frequency of configurations corresponding to the MWIS solution.

\section{QPU as a sampler}
\label{appendix:qpu_as_sampler}

\begin{figure}[h!]
  \includegraphics[width=\linewidth]{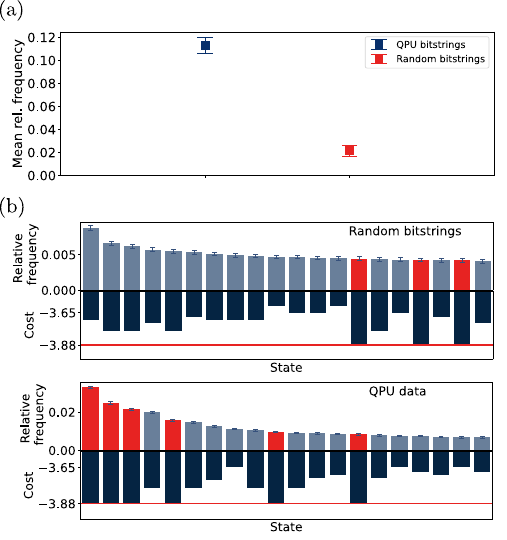} 
  \caption{Performance comparison between QPU-generated and
  randomly selected bitstring inputs.
  (a) Mean ground state relative frequencies computed over 100 independent 
  runs of classical post-processing, using 728 bitstrings obtained from
  the QPU and 728 randomly sampled bitstrings.
  (b) Aggregated histograms from 100 repetitions of the numerical experiment,
  showing the distribution of relative frequencies and costs for the most frequent states for random (top) and  QPU-derived bitstrings (bottom). 
  }
  \label{fig:qpu_vs_random}
\end{figure}
As a way to evaluate the QPU’s performance as a sampler despite noise, we compare its output to randomly generated bitstrings for the graph of \figref{fig:qubo_ata}(b). We collect 728 QPU-derived bitstrings from a fixed annealing ramp and generate an equal number of random bitstring samples. Each set is then used as input for 100 independent runs of the classical post-processing algorithm. As illustrated in \figref{fig:qpu_vs_random} (a), the QPU samples exhibit a notably higher mean ground state relative frequency. The QPU's advantage is reinforced by the histograms in \figref{fig:qpu_vs_random}  (b), which show a clear bias toward ground state configurations in the QPU data.

\newpage

%

\end{document}